\documentclass[%
 reprint,
superscriptaddress,
 amsmath,amssymb,
 aps,
]{revtex4-2}

\usepackage{graphicx}
\usepackage{dcolumn}
\usepackage{bm}

\usepackage[subrefformat=parens]{subcaption}

\newcommand{\z}{{\boldsymbol{z}}}

\newcommand{\x}{{\boldsymbol{x}}}
\newcommand{\w}{{\boldsymbol{w}}}
\newcommand{\1}{{\boldsymbol{1}}}

\begin{document}

\preprint{APS/123-QED}
\title{Continuous black-box optimization with quantum annealing and random subspace coding}

\author{Syun Izawa}
\affiliation{Department of Computational Biology and Medical Sciences, Graduate School of Frontier Sciences, The University of Tokyo, Kashiwa 277-8561, Japan}

\author{Koki Kitai}
\affiliation{Department of Mechanical Engineering, The University of Tokyo, Tokyo 113-8654, Japan}

\author{Shu Tanaka}
\affiliation{Department of Applied Physics and Physico-Informatics, Keio University, Yokohama 223-8522, Japan}
\affiliation{Green Computing System Research Organization, Waseda University, Tokyo, 162-0042, Japan}

\author{Ryo Tamura}
\email{tamura.ryo@nims.go.jp}
\affiliation{International Center for Materials Nanoarchitectonics, National Institute for Materials Science, Tsukuba 305-0047, Japan}
\affiliation{Research and Services Division of Materials Data and Integrated System, National Institute for Materials Science, Tsukuba 305-0047, Japan}
\affiliation{Department of Computational Biology and Medical Sciences, Graduate School of Frontier Sciences, The University of Tokyo, Kashiwa 277-8561, Japan}
\affiliation{RIKEN Center for Advanced Intelligence Project, Tokyo 103-0027, Japan}

\author{Koji Tsuda}
\email{tsuda@k.u-tokyo.ac.jp}
\affiliation{Department of Computational Biology and Medical Sciences, Graduate School of Frontier Sciences, The University of Tokyo, Kashiwa 277-8561, Japan}
\affiliation{RIKEN Center for Advanced Intelligence Project, Tokyo 103-0027, Japan}
\affiliation{Research and Services Division of Materials Data and Integrated System, National Institute for Materials Science, Tsukuba 305-0047, Japan}

\date{\today}

\begin{abstract}
A black-box optimization algorithm such as Bayesian optimization
finds extremum of an unknown function by alternating
inference of the underlying function
and optimization of an acquisition function. In a 
high-dimensional space, such algorithms perform poorly
due to the difficulty of acquisition function optimization.
Herein, we apply quantum annealing (QA) to overcome the
difficulty in the continuous black-box optimization.
As QA specializes in optimization of binary problems, a continuous vector has to be
encoded to binary, and the solution of QA has to be translated back.
Our method has the following three parts:
1) Random subspace coding based on axis-parallel hyperrectangles from continuous vector to binary vector.
2) A quadratic unconstrained binary optimization (QUBO) defined by acquisition function based on nonnegative-weighted linear regression model which is solved by QA.
3) A penalization scheme to ensure that the QA solution
can be translated back. 
It is shown in benchmark tests that
its performance using D-Wave Advantage$^{\rm TM}$ quantum annealer is competitive with a state-of-the-art method based on the Gaussian process in high-dimensional problems.  
Our method may open up a new possibility of quantum annealing and other QUBO solvers including quantum approximate optimization algorithm (QAOA) using a gated-quantum computers,
and expand its range of application to continuous-valued problems.  
\end{abstract}

\maketitle


\section{Introduction}
\begin{figure*}
  \begin{center}
    \begin{tabular}{c}
      \includegraphics[width=0.9\textwidth]{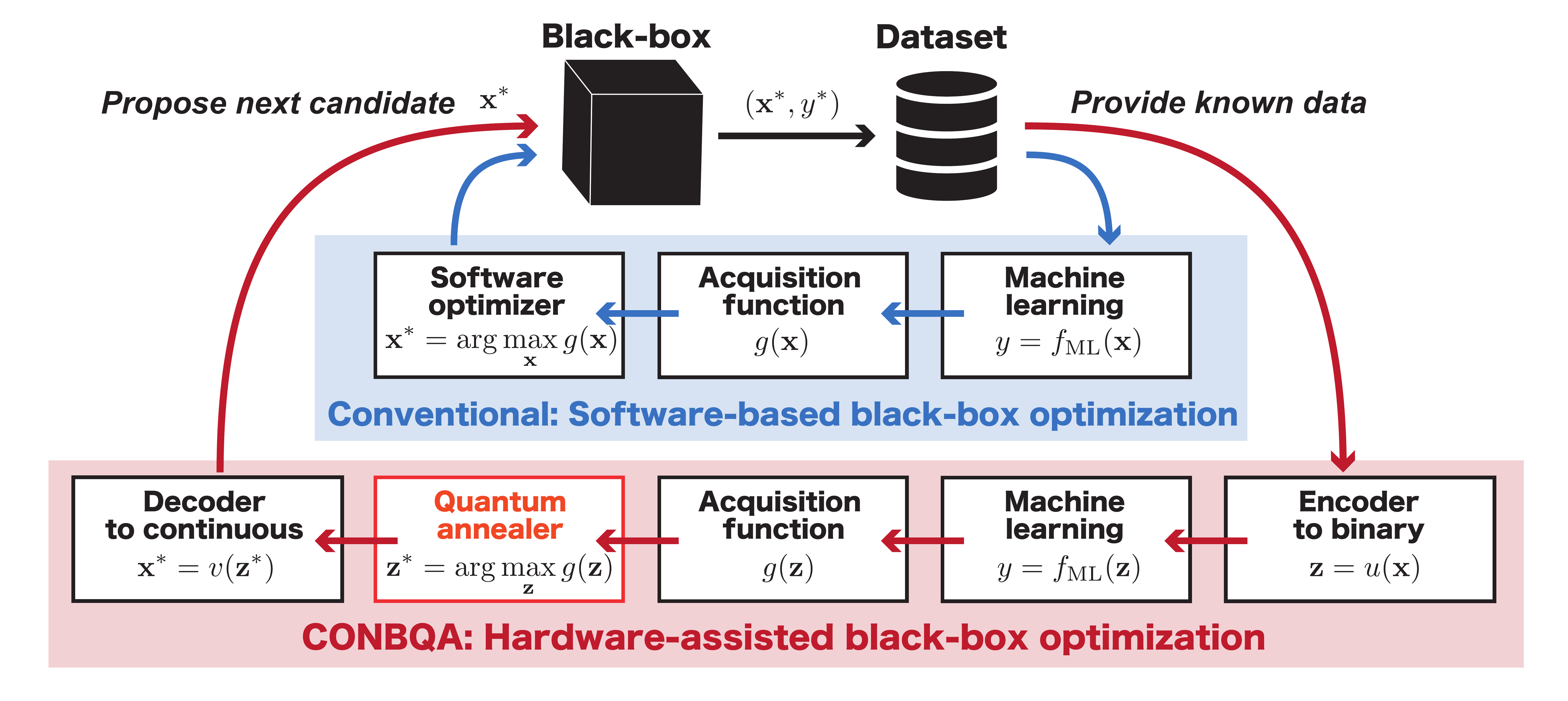}
    \end{tabular}
  \end{center}
  \caption{Software-based and hardware-assisted black-box optimization with continuous variables.
  To use special hardware, appropriate encoder and decoder between continuous and binary states should be put.}
\label{fig:hardware}
\end{figure*}

Given a black-box function $f(\x)$, $\x \in \Re^d$,
whose closed-form expression is unknown,
Black-box optimization (BO) identifies the extremum via an
iterative procedure of {\em input design} and {\em evaluation}~\cite{shahriari2015taking,terayama_2021}.
Figure~\ref{fig:hardware} shows the procedure of software-based black-box optimization.
First, the black-box function is estimated using a machine-learning prediction model $f_{\rm ML} (\x)$ trained by known dataset $D = \{ (\x_1,y_1),...,(\x_{n},y_{n}) \}$.
Using a trained model, an {\em acquisition function} $g(\x)$ is defined, and
the next input $\x^*$ is designed by optimizing it.
Next, the value of black-box function at the recommended point $y^* = f(\x^*)$ is
evaluated by a black-box system, e.g., simulation or experiment.
A new data $(\x^*, y^*)$ is added to $D$.
These steps are iterated as long as the budget allows.
Black-box optimization has been applied to many physics applications
including autonomous X-ray scattering experiments~\cite{noack2019kriging},
inverse scattering~\cite{vargas2019bayesian},
crystal structure prediction~\cite{yamashita2018crystal}, design of organic synthesis experiments~\cite{hase2018phoenics}, and effective model estimation~\cite{tamura2018plos}.

It is expected that BO accelerates extremum identification in comparison to random sampling.
Namely, at the same number of evaluations, the best input vector by BO
should be closer to the extremum than that of random sampling on average.
It is known that, however, the advantage of BO quickly diminishes as the
number of dimensionality increases~\cite{choffin2018scaling}.
There are two possible reasons, statistical and computational.
In statistical terms, it is difficult to estimate $f(\x)$ with a limited number of training data
in a high-dimensional space.
In computational terms, the optimization of a non-convex acquisition function becomes
increasingly difficult as the number of dimensionality increases. 

In this paper, to solve computational problem for continuous black-box optimization, we propose a hardware-assisted
black-box optimization algorithm called CONBQA (CONtinuous Black-box optimization with Quantum Annealing,
pronounced like kombucha) \cite{CONBQA}
that employs a quantum annealing\cite{kadowaki_quantum_1998,tanaka_quantum_2017} as an acquisition function optimizer.
Here, we use quantum annealers developed by D-Wave Systems, Inc.\cite{johnson_quantum_2011} which are special hardware for searching the ground state of Ising model or solving quadratic unconstrained binary optimization (QUBO).
For our method,
instead of quantum annealer,
we can use other Ising machines established using characteristic physics systems such as, digital circuit~\cite{matsubara_ising-model_2017}, CMOS~\cite{yamaoka_20k-spin_2015}, laser oscillator~\cite{inagaki_coherent_2016,mcmahon_fully_2016}, GPU~\cite{Goto-2019}, and so on.
Recently, Kitai et al.~\cite{kitai2020designing}
proposed a method that solves a discrete black-box optimization problem with the help of quantum annealing, which is called FMQA algorithm.
This method uses the factorization machine as a machine-learning prediction model, 
and can solve computational problem in discrete black-box optimization.
On the other hand, continuous black-box optimization cannot be performed by FMQA,
and technique to encode continuous data to binary data is required.
Note that some methods for discrete black-box optimization by Ising machines are recently developed~\cite{Hatakeyama_2021,koshikawa_2021}.

Figure~\ref{fig:hardware} illustrates the procedure of CONBQA. 
1) Continuous data $\x$ are encoded to binary data $\z$. 
2) Machine learning is performed and an acquisition function $g(\z)$ is prepared. 
3) The QUBO defined by $g(\z)$ and additional constraints to ensure decodability is solved by a quantum annealer to obtain maximizer $\z^*$. 
4) The solution is decoded back to a continuous vector $\x^*$. 
5) The value of black-box function $y^* = f(\x^*)$ is evaluated.
6) A new data point is added to the dataset. 

\begin{figure}[h]
  \begin{center}
    \includegraphics[width=0.5\textwidth]{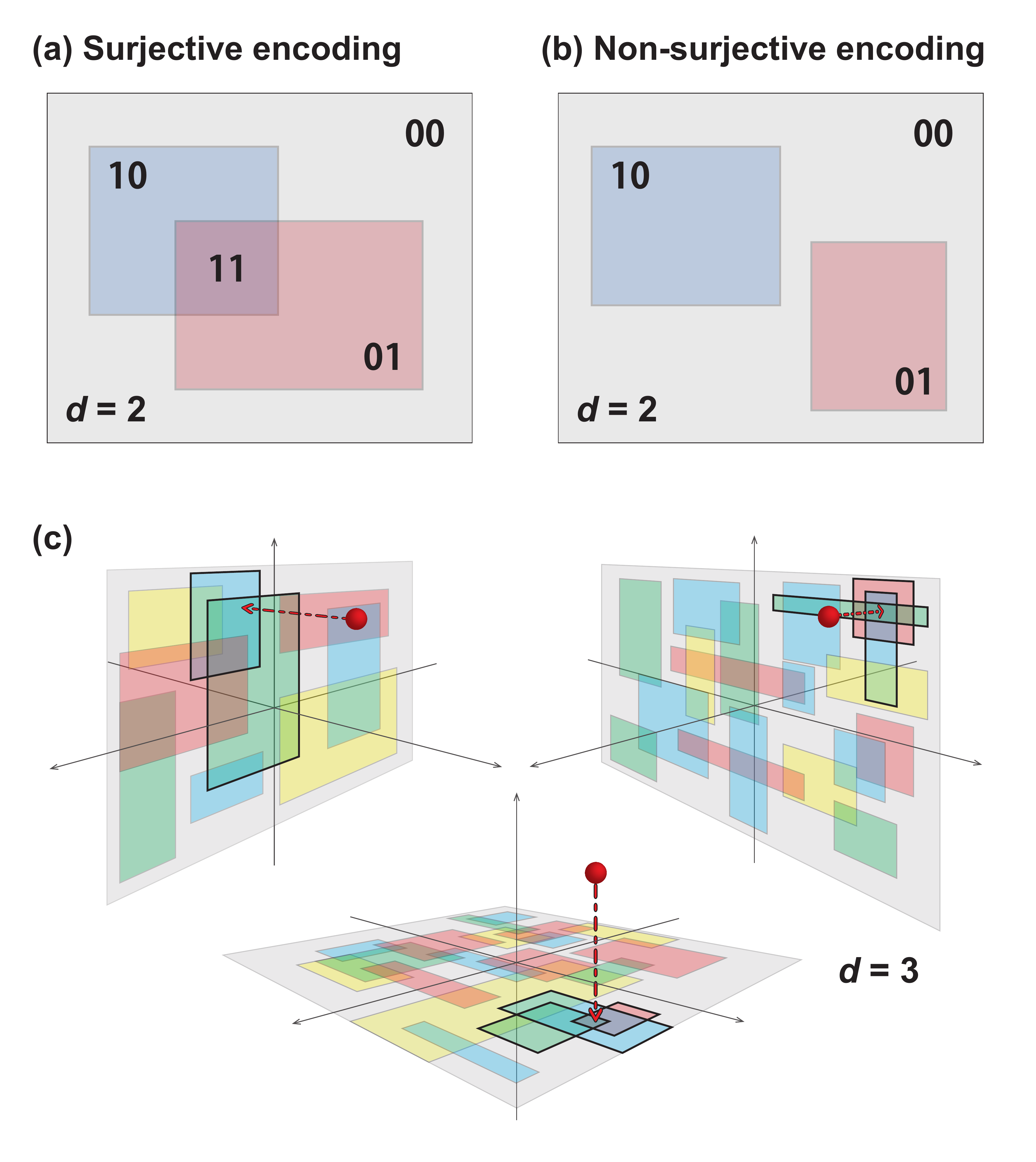}
  \end{center}
  \caption{
  (a) Surjective encoding for two-dimensional case when two-dimensional hyperrectangles are used.
If the continuous vector is included in the left blue hyperrectangle, the first bit is 1, while for the right red rectangle, second bit becomes 1 if continuous vector is located. 
(b) None-surjective encoding for two-dimensional case.
The vector `11' does not have any corresponding point.
(c) Random subspace coding in the three-dimensional continuous space.
  Although each square is drawn on each plane, it extends in the vertical axis for each place and forms rectangular.
  The bits expressing the bold rectangle are 1, and others 0 for the red point position.}
  \label{fig:rsc}
\end{figure}

Continuous-to-binary encoding is a common subject of research in scientific fields including
neuroscience~\cite{kanerva1988sparse,rachkovskii2005properties} and machine learning~\cite{eckstein2019repr,atzmueller2015subgroup}.
Among existing methods,
we employ random subspace coding~\cite{rachkovskii2005properties} due to its simplicity and high resolution. 
Several coordinates are randomly chosen and an axis parallel hyperrectangle is sampled randomly in the subspace.
A continuous vector is mapped to a binary variable by projecting it to the subspace.
If the mapped point is included in the hyperectangle, the vector is mapped to one, otherwise zero.
By using $m$ hyperrectangles, a continuous vector is mapped to an $m$-dimensional binary vector. 
In Fig.~\ref{fig:rsc} (a), example of random subspace coding in the two-dimensional continuous space is shown.
Here, two-dimensional coordinates are selected to generate hyperrectangles, and $m=2$ case is considered.
Two rectangles divide the whole space into four regions.
It corresponds to an encoding to a binary vector with two bits,
i.e., if the continuous vector is included in the left blue hyperrectangle, the first bit is 1, while for the right red rectangle, second bit becomes 1 if continuous vector is located. 
This example is surjective encoding.
On the other hand,
it is worth noting that the random subspace coding is not necessarily surjective.
Figure~\ref{fig:rsc} (b) is an example of non-surjective encoding.
In this case, there may exist binary vectors that cannot be decoded to continuous vectors, i.e., `11' in this figure.
In addition, for three-dimensional case,
an example of the random subspace coding is shown in Fig.~\ref{fig:rsc} (c).
Here, the continuous vector located at the red point is mapped to the binary vector where the bits expressing bold hyperrectangles are 1 and the others are 0.

In CONBQA, as the machine learning model in the binary space,
a nonnegative-weighted Bayesian linear model is employed,
and the acquisition function based on the trained model is defined.
We prove that the global optimal solution of the acquisition function under constraints to ensure decodability is always decodable via Helly's theorem.
Interestingly,
nonnegativity constraints in the prediction model are essential in proving the property.
To use quantum annealer, a QUBO is formulated with the acquisition function and additional constraints.
Note that it is not really necessary to obtain global optimal solution of QUBO by quantum annealing,
because decodable solutions are exist even if not global optimal solution.
Thus, the task of quantum annealing is to search decodable solutions with as large acquisition function as possible.
In numerical experiments using benchmark functions,
it was shown that our method is competitive with
Add-GP~\cite{kandasamy2015high}, which is one of the
state-of-the art method for high-dimensional Bayesian optimization~\cite{choffin2018scaling}.

\section{Method}

\subsection{Random Subspace Coding}
In random subspace coding in a $d$-dimensional space, the dimensionality of subspace $d_s$ is first determined.
Let ${\cal I} = \{i_1,\ldots,i_{d_s}\}$ denote the indices of randomly chosen coordinates from $[1,d]$,
and $\{[l_i, u_i]\}_{i \in {\cal I}}$ be closed intervals in respective coordinates.
Here, $l_i < u_i$, $l_i \in \Re$ and $u_i \in \Re$.
Random sampling of each interval $[l_i, u_i]$ is performed using the Dirichlet process method~\cite{devroye1993generating}.
Given a positive integer $N$, this method can generate an interval such that
any point in $[0,1]$ along each selected coordinate is included in the interval with probability $1/N$.
Throughout this paper, $N$ is fixed as $3$.
Then, an axis-parallel hyperrectangle $R$ is determined as 
\begin{eqnarray}
R = \{ \x \in \Re^d \; | \ \;  l_i \le x_i \le u_i, i \in {\cal I} \}.
\end{eqnarray}
Later on, we denote it as ``rectangle'' for simplicity.  
Using $m$ rectangles $R_1,\ldots,R_m$,
the random subspace coding is defined as $z_k = I(\x \in R_k)$,
where $I(\cdot)$ is one if the condition is satified and zero otherwise.

Given a binary vector $\z = \{ z_k \}_{k=1,...,m}$, a point $\x \in \Re^d$ corresponds to $\z$,
if $\x$ is included in all rectangles with $z_k = 1$ and not
included in those with $z_k=0$.
Let $P$ denotes the intersection of all rectangles with $z_k = 1$,
\begin{eqnarray}
P(\z) = \bigcap_{\{ k \; | \; z_k = 1 \} } R_k. 
\end{eqnarray}
Let $N$ denote the union of all rectangles with $z_k = 0$,
\begin{eqnarray}
N(\z) = \bigcup_{\{ k \; |  \; z_k = 0 \} } R_k.
\end{eqnarray}
Since the preimage of $\z$ is described as $P(\z)-N(\z)$,
$\z$ is {\em decodable} if and only if $P(\z) -N(\z) \neq \emptyset$.

\subsection{Black-Box Optimization}
In CONBQA, the training dataset $D$ are encoded to 
binary dataset $D^\prime = \{(\z_1,y_1),...,(\z_n,y_n) \}$ through random subspace coding.
Here, $\{ y_i \}_{i=1,...,n}$ is normalized by min-max normalization,
and thus these values are nonnegative.
Then, a QUBO is prepared from $D^\prime$ and
its optimal solution $\z^*$ is obtained with a quantum annealing.
Next, $\z^*$ decoded to a continuous vector $\x^*$, and
fed to the black-box system.

To model an underlying function from $D^\prime$, a nonnegative linear model: 
\begin{equation} \label{eq:logic}
y = \sum_{k=1}^m w_k z_k, \; w_k \ge 0,
\end{equation}
is employed, and trained with least-squares fitting to $D^\prime$.
Let $\w^* = \{ w_k^* \}_{k=1,...,m}$ denote the trained parameter vector.
Thus, the acquisition function is defined as
\begin{equation}
g(\z) = \sum_{k=1}^m w_k^* z_k. \label{eq:acq_binary}
\end{equation}
The unique solution with maximum of $g(\z)$ has $z_k = 1$ for $\forall k$.
However, this solution may be a non-decodable solution,
because it is almost impossible to exist the region, where all rectangles are intersected.
That is, unconstrained maximization of the acquisition
function may lead to a non-decodable solution.
To keep the optimal solution decodable,
we add the following constraint, $\z \in C$ where
\begin{equation} \label{eq:noedge}
  C = \{ \z \in \{0,1\}^m \; | \; z_i z_j = 0, \; {\rm if} \; R_i \cap R_j = \emptyset \}.  
\end{equation}
The meaning of this constraint is that the data point do not belong to $R_i$ and $R_j$ when two rectangles are not intersected.
The constrained problem is then described as
\begin{equation} \label{eq:thompson}
\max_{\z \in C} \sum_{k=1}^m w^*_k z_k. 
\end{equation}
In Sec. III, we will present a proof that
the optimal solution of Eq.~(\ref{eq:thompson}) is decodable.

The problem by Eq.~(\ref{eq:thompson}) is treated as a QUBO and
solved by a quantum annealing.
As done in many application studies of quantum annealing,
constraints are replaced by a penality term as follows,
\begin{equation} \label{eq:qubomwc}
\min_{\z \in \{ 0,1 \}^m}
  - A  \sum_{i = 1}^m w_i^* z_i + B \sum_{ \{ ( i,j ) | R_i \cap R_j = \emptyset, i<j \}} z_i z_j,
\end{equation}
where $A$ and $B$ are positive hyperparameters.
To make contributions from two terms comparable,
we use $A=1/\max_k \{ w_k^* \}$ and $B=1$.
Once the solution $\z^*$ of Eq.~(\ref{eq:qubomwc}) is obtained by quantum annealing,
a point $\x^*$ is randomly sampled from the preimage $P(\z^*)-N(\z^*)$,
evaluated to yield a new outcome $y^*$ by the black-box system. 
The new data point $(\x^*,y^*)$
is then added to the dataset $D$, and the next iteration is put forward.

\section{Decodability}
To understand Eq.~(\ref{eq:thompson}) better,
we formulate it as a graph-theoretic problem.
Let us define $m$ nodes corresponding to all rectangles,
and make an edge if the corresponding rectangles are overlapped. 
Put a nonnegative weight $w^*_k$ to each node. 
Then, the acquisition function defined by Eq.~(\ref{eq:acq_binary}) is the sum of all node weights
of the subgraph induced by the nodes with $z_k = 1$.
Let us call the rectangles with $z_k =1$ and $0$ positive
and negative rectangles, respectively.
The nodes corresponding to positive and negative rectangles
are called positive and negative nodes, respectively.

Consider the following two conditions.
1) There is no edge between node $i$ and $j$.
2) Both of the nodes are positive.
If both of the conditions are satisfied for $\z$, it is not decodable,
because the intersection of two non-overlapping rectangles is empty,
hence $P(\z)=\emptyset$.
The constraint of Eq.~(\ref{eq:noedge}) ensures that this situation does not
happen, and the positive node set always forms a clique.
Assume that the positive set does not form a clique. Then, there is at least
one pair of positive nodes without an edge, violating the constraint.
In summary, the optimization problem corresponds to a maximum clique
problem~\cite{chapuis2019finding}, 
where the weight of a clique is defined as the sum of
all node weights.

Let us prove that the global optimal solution
$\z^*$ of Eq.~(\ref{eq:thompson}) is decodable. As shown before,
the decodability condition is written as $P(\z^*)-N(\z^*) \neq \emptyset$.
To prove it, it is sufficient to show $P(\z^*) \neq \emptyset$
and $P(\z^*) \cap N(\z^*) = \emptyset$.
Since the optimal solution corresponds to a clique in the graph theoretic problem,
all pairs of positive rectangles overlap.
Due to Helly's Theorem of convex sets, 
the intersection of all rectangles $P(\z^*)$ is nonempty.
Let us assume that $P(\z^*) \cap N(\z^*) \neq \emptyset$.
Then, $P(\z^*)$ has to have overlap with one or more negative rectangles.
If a negative rectangle has overlap with $P(\z^*)$,
it has overlap with all the positive rectangles.
In that case,
one can build a larger clique than the current one by including the
negative rectangle. 
Since node weight is positive,
larger clique should have larger weight.
It contradicts the assumption that 
the current clique has maximum weight, thus $P(\z^*) \cap N(\z^*) = \emptyset$.

\section{Experiments}
\subsection{Results with D-Wave Quantum Annealer}
Using D-Wave Advantage, we implement a hardware-assisted system for continuous black-box
optimization. 
In the previous section, we proved that, if the hardware optimizer
finds the optimal solution of Eq.~(\ref{eq:thompson}), it is always decodable. In reality, however,
a quantum annealer and other algorithms solve the QUBO problem defined by Eq.~(\ref{eq:qubomwc}) and do not always return the ground state.
Thus, there are the following three possible cases when the solution $\z^*$ is calculated by quantum annealing:
\begin{enumerate}
\item Empty: $P(\z^*) = \emptyset$.
\item Admissible: $P(\z^*) \neq \emptyset$ and $P(\z^*) \cap N(\z^*) \neq \emptyset$.
\item Decodable: $P(\z^*) \neq \emptyset$ and $P(\z^*) \cap N(\z^*) = \emptyset$.
\end{enumerate}
For a decoder,
if $\z^*$ with decodable or admissible is recommended,
the next data point $\x^*$ is randomly sampled from the region with $P(\z^*) \neq \emptyset$.
On the other hand,
when the solution is empty,
$\x^*$ is randomly selected from the whole search space.

\begin{figure}
  \begin{center}
    \begin{tabular}{c}
      \includegraphics[width=0.45\textwidth]{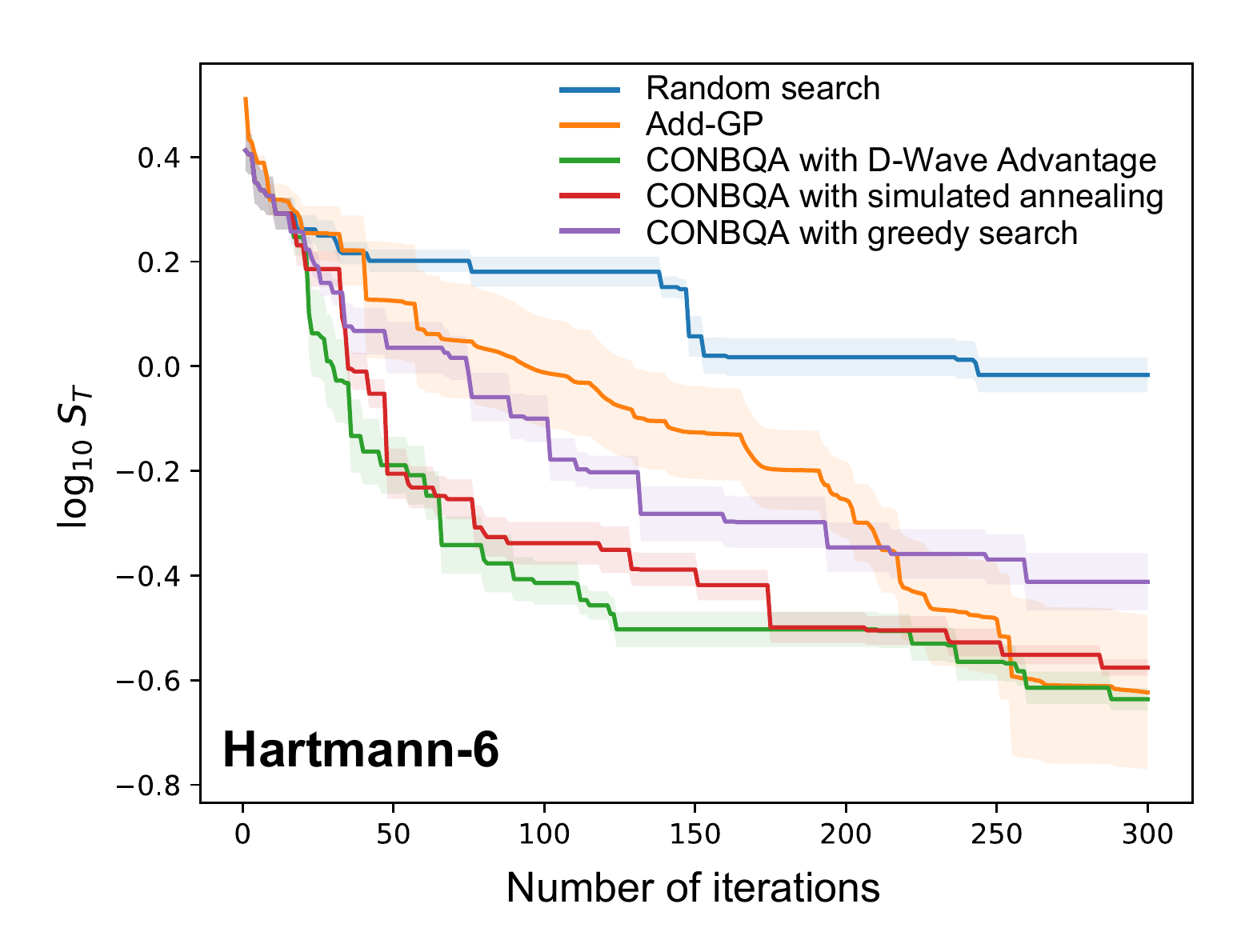}
    \end{tabular}
  \end{center}
  \caption{Regret $S_T$ depending on the number of iterations by random search, Add-GP, and CONBQA. CONBQA with D-Wave Advantage with hybrid solver service performs better than CONBQA with simulated annealing and that with greedy search. It outperforms Add-GP based on UCB score and random search as well. 10 independent runs are performed and shaded region is the error.}
\label{fig:hartmann}
\end{figure}

To test the performance, we employ
a 6-dimensional function called Hartmann-6~\cite{picheny2013benchmark} that is
often used for testing global optimizers.
In this testing function, whole search spaces are determined as $[0,1]$ in all coordinates.
For all experiments, the parameters of random subspace coding are determined
as $m=60$ and $d_s=2$.
The performance of CONBQA is measured by regret $S_T$,
which is the difference between the optimal value of Hartmann-6 and the best one observed so far in the black-box optimization.
Thus, the purpose of the present continuous black-box optimization is to obtain small $S_T$ with as few iterations as possible.
Figure~\ref{fig:hartmann} shows the regret in logarithmic scale
against the number of iterations for CONBQA, AddGP based on UCB score, and random search.
In the 15 initial steps,
$\x$ is randomly sampled from the whole search space, and afterwords, iterations controlled by CONBQA or Add-GP are performed.
Here, the average over 10 independent runs, where initial randomly-sampled data are different, is shown with the error indicating the standard deviation.

As a QUBO solver used in CONBQA, we employ D-Wave Advantage with hybrid solver service\cite{D-Wave_HSS}, simulated annealing using dwave-neal with default setting\cite{neal},
and greedy search.
It is remarkable that when the number of iterations is small, 
regardless of QUBO solvers,
CONBQA outperforms Add-GP which performed best
in a previous benchmark study~\cite{choffin2018scaling}.
Of course, in comparison with random search,
better optimization performance of CONBQA is confirmed.
This result indicates that our framework to solve continuous black-box optimization using random subspace coding and QUBO solvers as acquisition function optimizer is work well.
In addition,
among the CONBQA variants, the quantum annealer performed best.
In particular, optimization result by quantum annealer is quite better than that by greedy search, indicating that the performance of continuous black-box optimization can possibly be improved by a better QUBO solvers, i.e., specialized hardware.

Next, we address the decordability in optimization process.
Table~\ref{tab:hartmann} shows the fraction of empty, admissible, and decodable solutions in the iterations in CONBQA depending on the QUBO solvers.
As an important fact, we have never observed empty solutions by three methods to solve QUBO.
Simulated annealing and greedy search had a
higher fraction of decodable solutions than the quantum annealer
despite that they performed poorly in optimization (see Fig.~\ref{fig:hartmann}).
This result indicates that the solutions obtained from the simulated annealing or greedy search are well satisfying the constraint of Eq.~(\ref{eq:noedge}),
but tend to stuck at local minima with larger values of the first term in Eq.~(\ref{eq:qubomwc}).
On the other hand, quantum annealer recommends many admissible solutions, which were not fine-tuned for the constraint, but better optimization is realized.
This means that the solutions by quantum annealer have smaller value of the first term in Eq.~(\ref{eq:qubomwc}) instead of fulfilling the constraint exactly,
and CONBQA could get closer to the global optimal solution.

\begin{table}[]
    \centering
    \caption{Fraction of empty, admissible, and decodable solutions obtained from three different QUBO solvers.}
    \begin{tabular}{c|ccc}
         \hline\hline
         QUBO solver & \ Empty \ & \ Admissible \ & \ Decodable \ \\
         \hline
         D-Wave Advantage & 0 \% & 53.5 \% & 46.5 \%   \\
         Simulated annealing & 0 \% & 12.1 \% & 87.9 \% \\
         Greedy search & 0 \% & 3.6 \% & 96.4 \% \\
         \hline\hline
    \end{tabular}
    \label{tab:hartmann}
\end{table}

\subsection{Benchmarking with Simulated Annealing}
\begin{figure}
  \centering
  \begin{tabular}{c}
  \includegraphics[width=0.4\textwidth]{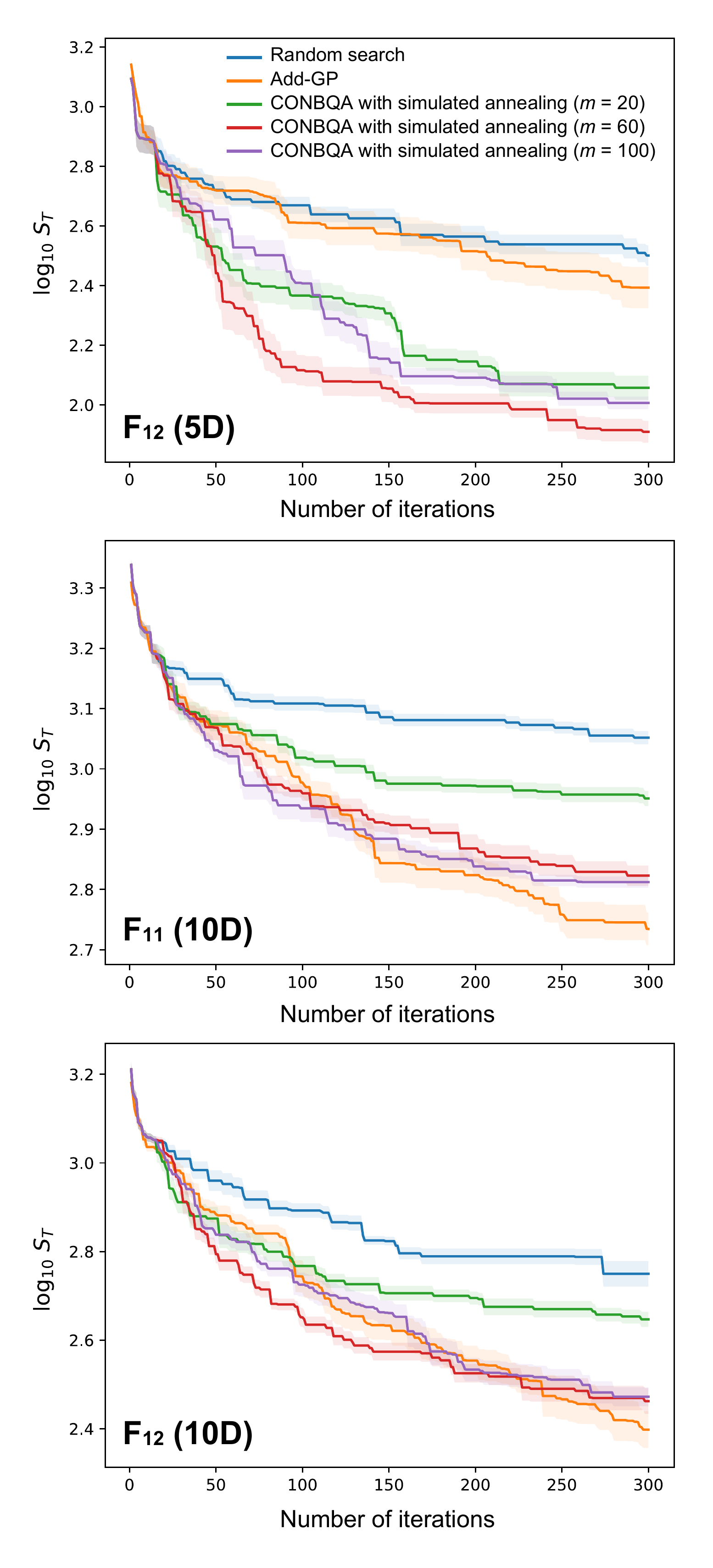}
  \end{tabular}
  \caption{Performance of CONBQA with 20, 60, and 100 bits in comparison to random search and AddGP with respect to three benchmark functions ($F_{12}$ (5D), $F_{11}$ (10D), and $F_{12}$ (10D))~\cite{Li_2013}.
  Simulated annealing was used to solve QUBO.
  10 independent runs are performed and shaded region is the error.}
  \label{fig:benchmark}
\end{figure}

The performance of CONBQA is tested further with three additional benchmark functions
from 2013 IEEE CEC Competition on Niching Methods for Multimodal Optimization~\cite{Li_2013}.
Among the 20 functions provided, three high-dimensional
functions of $F_{12}$ (5D), $F_{11}$ (10D), and $F_{12}$ (10D) are chosen.
Their dimensionality is 5, 10, and 10, respectively.
To try some experimental settings with different number of encoding bits, $m$,
we perform CONBQA with simulated annealing. 
In these testing functions, whole search space are determined as $[0,1]$ in all coordinates.
Figure~\ref{fig:benchmark} shows the regret in logarithmic scale against the number of iterations for CONBQA with $m=20, 60$, and $100$,
Add-GP, and random search.
For random subspace coding,
$d_s =2$ is used.
The average over 10 independent runs, where initial 15 randomly-sampled data are different, is shown with the error indicating the standard deviation.

For relatively low-dimensional dataset of $F_{12}$ (5D),
CONBQA is superior in finding better solutions in comparison with Add-GP
regardless of $m$.
In this case,
$m=60$ shows the better optimization performance.
If $m$ is increased, the resolution of random subspace coding becomes high, which is discussed in the next subsection, 
and we can get closer to the optimal solution in principle.
On the other hand, the regression model becomes complicated,
and in an early stage of iterations,
an accurate prediction model is difficult to construct by statistical problems.
Thus, an appropriate value of $m$ will be determined by this trade-off
and is strongly depending on the target problem.

For 10-dimensional datasets ($F_{11}$ (10D) and $F_{12}$ (10D)), 
CONBQA outperforms Add-GP in an early stage, but slightly worse than Add-GP
at the final stage of optimization.
This is because CONBQA cannot get exactly to the optimal point 
due to the finite resolution of random subspace coding.
Note that this resolution problem can be resolved by using large $m$,
but the trade-off described above should be appeared.
Thus, it is not always necessary to make $m$ larger to obtain better optimizations.
In applications such as experimental and simulation conditions tuning in physics, chemistry, and materials science,
the number of iterations is severely limited.
That is, the early stage of optimization is more important for these applications.
In such cases, CONBQA would be preferred over Add-GP.

\subsection{Higher Dimensionality}
In the benchmark, CONBQA was applied to up to 10 dimensional problems. 
Since the ability of Ising machines is growing rapidly,
CONBQA will likely be applied to problems with higher dimensionality in the future. 
To estimate how many bits are necessary for CONBQA, we investigated the resolution of
random subspace coding for $d=10,100$, and $1000$.
The resolution is characterized by the average size of
the intersection of all rectangles enclosing a randomly chosen point which is represented by $R_I$.
The size of a rectangle is measured by the average of its edge lengths.
If $R_I$ is small, in a decoder,
the possible region of $\x^*$ is small when $\z^*$ is given.
Thus, the high resolution by random subspace coding is realized when $R_I$ is small enough.

\begin{figure}
  \begin{tabular}{c}
  \includegraphics[width=0.4\textwidth]{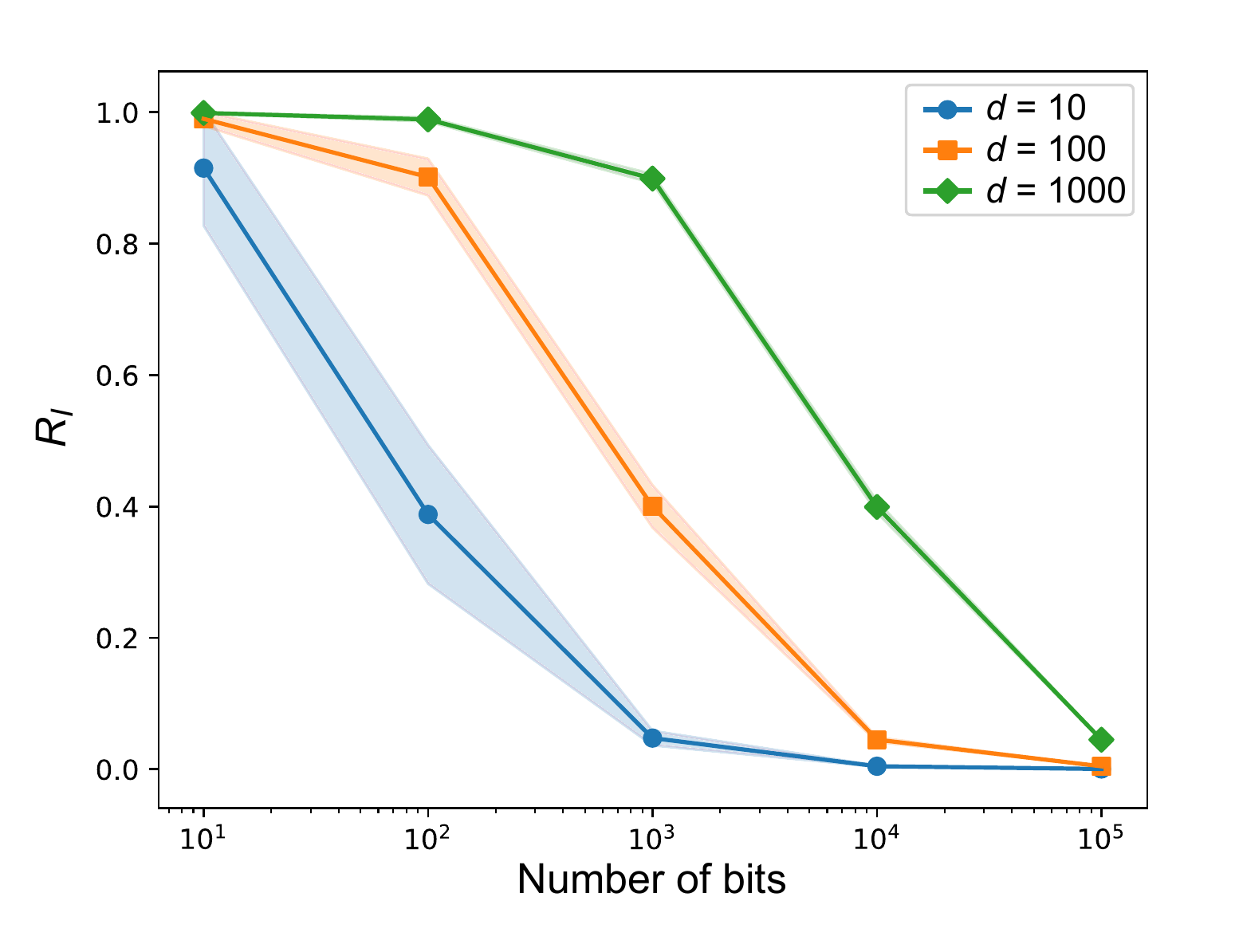}
  \end{tabular}
\caption{Average size of the intersection of all rectangles enclosing a randomly chosen point for random subspace coding $R_I$ depending on the number of bits. Results at various dimensionality of problem $d$ are shown. The number of randomly chosen points is 50 and shaded region is the error.} 
\label{fig:resolution}
\end{figure}

Figure~\ref{fig:resolution} shows $R_I$ depending on the number of bits for $d=10, 100$, and $1000$ for $d_s = 2$.
Here, the average over 50 randomly chosen points is shown with the error indicating the standard deviation.
In our benchmarking in Sec. IV B,
if 100 bits are used for 10-dimensional problems,
a better performance than Add-GP in an early stage of optimization is realized in CONBQA.
According to this fact,
$R_I \simeq 0.4$ would be enough to work CONBQA effectively for $d=10$.
Thus, with 10,000 bits, it is possible to reach sufficiently
high resolution for 1,000-dimensional problems.
For 100-dimensional problems, 1,000 bits will be sufficient.
On the other hand, 
to get more closer to the optimal solution at the final stage of optimization,
it is important to improve a resolution.
In this case, we will need ten times more bits for each dimension of problems at least.
Notice that statistical aspects are totally disregarded in this analysis:
solving a high-dimensional problem may require a
prohibitive number of black-box evaluations.
Then, the required number of bits is strongly depending on target problems.
Nevertheless, CONBQA has a potential to solve high-dimensional continuous black-box optimization problems
that are difficult to solve by conventional methodologies when more bits will be used in quantum annealer.

\section{Conclusion}
To summarize, we presented that a quantum annealing can be used to solve
continuous black-box optimization problems,
if a coding-decoding scheme and a machine learning algorithm are properly designed and combined.
In this study, for a coding-decoding scheme, random subspace coding was applied.
Mathematical foundation about decodability was laid out with
intriguing connection to Helly's Theorem.
It was shown that CONBQA is competitive with a
state-of-the-art black-box optimization algorithm,
encouraging applications to various domains.
In addition, random subspace coding will be one of the effective methods to handle continuous problems by Ising machines.

CONBQA is the first of its kind and there remains many
possibilities for improvement.
First, uncertainty quantification~\cite{smith2013uncertainty} can be combined with CONBQA.
In black-box optimization, it is customary to compute uncertainty of
prediction and incorporate it in the aquisition function,
while CONBQA does not have such a mechanism.
A special aspect about CONBQA is that optimization of the acquisition function
is done by a quantum annealer, hence the solution is affected by noise.
It is interesting to investigate how
uncertainty quantification and quantum noise affect overall performance.
Second, CONBQA currently make one suggestion at a time to the black-box system,
but, in many cases, it is more desirable to make multiple suggestions simultanously.
A quantum annealer is basically a sampler and generates many solutions at a time.
Currently, only the best solution is used,
but it may be possible to use other solutions as multiple suggestions.

In our algorithm,
not only quantum annealer but also all QUBO solvers including quantum approximate optimization algorithm (QAOA)~\cite{Farhi_2014,Hadfield_2017,Otterbach_2017} using a gated-quantum computers are adopted.
Thus, CONBQA may open up a new possibility of these machines to continuous valued-problems.

\begin{acknowledgments}
This work is supported by AMED under Grant Number JP20nk0101111,
the New Energy and Industrial Technology Development Organization (NEDO) (Grant No. P15009), 
Council for Science, Technology and Innovation (CSTI), Cross-ministerial Strategic Innovation Promotion Program (SIP) (Technologies for Smart Bio-industry and Agriculture, “Materials Integration” for Revolutionary Design System of Structural Materials, and Photonics and Quantum Technology for Society 5.0), 
and JST ERATO (Grant No. MJER1903).
\end{acknowledgments}

\providecommand{\noopsort}[1]{}\providecommand{\singleletter}[1]{#1}%
%


\end{document}